\documentclass[12pt, letterpaper]{article}
\pdfoutput=1
\usepackage[left=25.4mm,right=25.4mm,top=25.4mm,bottom=25.4mm]{geometry}
\usepackage[utf8]{inputenc}
\usepackage[version=3]{mhchem} 
\usepackage{siunitx}
\usepackage{graphicx}
\usepackage{amsmath}
\usepackage{authblk}
\usepackage{subcaption}
\usepackage{soul}
\usepackage{color}
\usepackage{float}
\usepackage[table]{xcolor}
\usepackage{xltabular}
\begin{document}

\title{Turbulent flame speed based on the mass flow rate: theory and DNS}
\author[1]{Swetaprovo Chaudhuri \thanks{Corresponding author, email: swetaprovo.chaudhuri@utoronto.ca}}
\author[2]{Bruno Savard \thanks{email: bruno.savard@polymtl.ca}}

\date{}

\affil[1]{Institute for Aerospace Studies, University of Toronto, Toronto, Canada}
\affil[2]{Department of Mechanical Engineering, Polytechnique Montréal, Montréal, Canada}

\maketitle

\begin{abstract}
Starting with an integral formulation of mass flow rate through an ensemble of isotherms constituting a statistically planar, turbulent premixed flame, a scaling for the corresponding turbulent flame speed is derived without invoking Damk{\"o}hler's hypotheses. Major approximations and interim results are validated using a large Karlovitz number, unity Lewis number, Direct Numerical Simulation (DNS) dataset of n-heptane/air mixture, computed with reduced chemistry.  A new length scale quantifying the fluctuation distance of the isotherms within the premixed flame structure is introduced.
\end{abstract}


\section{Introduction}
Turbulent flame speed has been a topic of active research since Damk{\"o}hler's  1940 paper \cite{damkohler1940einfluss}. Of the two hypotheses presented in that paper, the first hypothesis suggests that for premixed flames in large-scale turbulence, the turbulent flame speed is proportional to the flame surface area. The second hypothesis suggests that for premixed flames in intense, small-scale turbulence, the turbulent diffusivity instead of thermal diffusivity determines the turbulent flame speed, which results in a corresponding $Re_T ^{1/2}$ scaling. Here $Re_T = u_{r m s }^{\prime} l_I/(S_L l_F)$ and $S_L$ is the flame speed of an unstretched, planar, laminar premixed flame \cite{law2006}. $u_{r m s }^{\prime}$ is the root mean square of the fluctuating velocity. $l_I$ is the integral length scale, $l_F$ is the diffusion thickness of the flame. Note $S_L l_F=\alpha$ the thermal diffusivity of the unburned reactants, and with unity Schmidt number $Sc=1$ assumption $\nu = S_L l_F$. The reader is recommended to refer to Driscoll et al. \cite{driscoll2020} for a state of the art interpretation of Damk{\"o}hler's  hypotheses based on recent measurements. Damk{\"o}hler directly extrapolated the result: $S_L \sim \sqrt{\alpha w_b^0}$, to turbulent flame speed $S_t \sim \sqrt{\alpha_T w_b^0}$, with the assumption that the reaction rate of the standard planar laminar premixed flame $w_b^0$, and hence the chemical time scale, remains invariant between the corresponding laminar and turbulent configurations \cite{law2006, Peters2000}. Defining turbulent diffusivity $\alpha_T = u_{r m s }^{\prime} l_I$ leads to $S_t/S_L = Re_T ^{1/2}$. While the approach is elegantly simple, the difficulty lies in arriving at it systematically once a proper, turbulent premixed flame configuration is defined. Both turbulent flame speed and turbulent diffusivity are statistical quantities. As such, in turbulent combustion, turbulent diffusivity is applicable to an ``averaged" flame structure. In such an averaged flame structure, the corresponding diffusion length scale, average width of the reaction zone, and averaged reaction rates need not be equal to the standard laminar flame preheat zone thickness, reaction zone thickness, and $w_b^0$, respectively, though such equivalences are necessary to systematically arrive at the scaling, using Damk{\"o}hler's approach.

Since 1940s several expressions of the form $S_t = f(S_L, u_{r m s }^{\prime}, l_I, l_F)$ have been sought. Clavin and Williams suggested a quadratic relation between $S_t$ and $u_{r m s }^{\prime}$ \cite{clavin1979theory} for weak turbulence. Yakhot used renormalization theory to derive an expression for $S_t$ \cite{yakhot1988propagation}. Note that here, the symbol $S_t$ is used to refer to turbulent flame speed, in general. A specific definition of turbulent flame speed with a different symbol will be taken up in the following section. Lipatnikov and Chomiak \cite{lipatnikov2007global} introduced a turbulent Markstein number to obtain a turbulent flame speed expression for expanding turbulent flames. Kolla et. al \cite{kolla2009scalar} used the transport equation of the averaged scalar dissipation rate, their associated closures, and the KPP theorem \cite{lipatnikov2002turbulent} to derive an expression for turbulent flame speed for large Damk{\"o}hler number ($Da$) flames. Kerstein et al. \cite{kerstein1988} used the $G-$equation to derive an expression for $S_t/S_L$ as a volume integral of the absolute gradient of the level set function $G$ \cite{ kerstein1988}. The $G-$equation approach was also utilized by Peters \cite{Peters2000} to obtain turbulent flame speeds in both the corrugated and thickened flamelet regimes. Chaudhuri et al. \cite{chaudhuri2011} extended Peters' spectral formulation of the $G-$equation \cite{peters1992spectral} to derive an expression for turbulent flame speed. Some of the issues with the $G-$equation approach are restrictive initial condition as in $\cite{kerstein1988}$, averaging problems $\cite{sabel2010rigorous, oberlack2001symmetries}$, and their interpretation towards obtaining scaling relations for the dissipation rate of the level-set variable itself. The difficulty in utilizing averaged transport equations of progress variables to determine turbulent flame speed involves closures of the several terms and their applicability over large ranges of $Re_T$ and $Ka$. Here $Ka$ is the Karlovitz number. The reader could refer to the following review papers and books for a much more detailed exposition on this topic \cite{gulder1991turbulent,lipatnikov2002turbulent,lipatnikov2005molecular, driscoll2008turbulent, poinsot2005theoretical, driscoll2020, lipatnikov2012fundamentals, lieuwen2021unsteady, swaminathan2011turbulent}.

In many instances, the derivations of turbulent flame speed begin with Damk{\"o}hler's hypotheses and often the turbulent flame speed itself is not rigorously defined. Perhaps there is a need to revisit turbulent flame speed theoretically, from first principles, without invoking any hypothesis. In this paper, we initiate the derivation of turbulent flame speed of statistically planar premixed flames, from first principles, by rigorously estimating the averaged mass flow rate through an ensemble of propagating iso-scalar surfaces constituting a turbulent premixed flame. Such a mass flow rate underpins the turbulent flame speed defined in this paper. However, in later stages, we do utilize closure of local flame displacement speed and mean scalar dissipation rate. The major assumptions regarding interim closures, as well as interim results are validated with a high Karlovitz number, unity Lewis number, n-heptane/air DNS dataset by Savard et al. \cite{savard2015structure}. All figures presented in the present paper are obtained from this particular DNS dataset.

\section{DNS dataset}

The configuration consists of a statistically planar premixed turbulent flame in a cuboid domain, with inflow-outflow boundary conditions in the axial ($x$-) direction, and periodic boundary condition in the transverse ($y$- and $z$-) directions. The dimensions are $11L$, $L$, and $L$ in the $x$-, $y$-, and $z$-directions, respectively, with $L=5.3l_I$. 
An unburnt mixture of n-heptane and air with equivalence ratio of 0.9 and temperature of 298 K is specified. The pressure is uniform and constant at 1 atm. The ratio of integral scale to flame thermal thickness, $\delta_L$, is $l_I/\delta_L=1.0$, and $u'_{rms}/S_L=21$. The flame falls nominally at the limit between the thin reaction zones regime and the broken/distributed reaction zones regime, with Karlovitz number $Ka\approx\left[\left(u'_{rms}/S_L\right)^3/\left(l_I/l_F\right)\right]^{1/2}=96$. 

The inflow is generated from a separate homogenous isotropic turbulence simulation. Since no turbulence generation mechanism (e.g., through shear
layers or near wall) is present, in the absence of forcing, turbulence injected at the inlet would decay over a time scale given by $l_I/u'_{rms}$, which is here significantly shorter than the flame
characteristic time. To preserve turbulence across the domain,
the linear velocity forcing method is used~\cite{rosales2005linear,carroll2013proposed}. This forcing technique could interfere with
potential effects of the flame on the turbulence (e.g., turbulence
generation through pressure-dilatation). However, for flames with
moderate-to-high Karlovitz number as in the present study, this
effect can be neglected compared to viscous dissipation~\cite{bobbitt2016vorticity,wang2016turbulence,macart2018effects}.
This result was consistently obtained by Bobbitt et al.~\cite{bobbitt2016vorticity}, who
used the same configuration as in the present study, and Wang
et al.~\cite{wang2016turbulence} and MacArt et al.~\cite{macart2018effects}, who did not use turbulence forcing in their simulation where turbulence was generated through strong
mean shear, in round or slot jet configurations.

The governing equations for low-Mach number reactive flows are solved numerically with the finite difference, energy conserving solver NGA~\cite{desjardins2008high}. The scheme selected is second-order accurate in both space
and time. A semi-implicit Crank-Nicolson time integration
scheme is used~\cite{pierce2001progress,savard2015computationally}. The third-order Bounded QUICK scheme,
BQUICK~\cite{herrmann2006flux}, is used as the scalar transport scheme. The species diffusivities are set equal to the thermal diffusivity to enforce unity Lewis numbers. The chemical kinetics are described by a 35-species and 217-elementary reaction  mechanism~\cite{savard2015broken} reduced from~\cite{BISETTI2012}.

Data is collected over 30 integral eddy turnover times, during which the flame is in a statistically steady state. More details about the DNS dataset and the numerical approach can be found in~\cite{savard2015structure,savard2015broken}.

\section{Analysis}
Consider a statistically planar, statistically stationary, turbulent premixed flame in a rectangular cuboid domain with a square cross-section, with inflow-outflow boundary conditions in the axial directions, and with periodic boundaries in the transverse directions, as in the validation dataset described above. Turbulent, premixed reactants enter from the square-faced inlet with a mean velocity in the $x-$direction to interact with the statistically stationary premixed flame downstream. Products leave the squared-faced outlet. We are only interested in conditions with turbulence Reynolds number $Re_t=u^{\prime}_{rms}l_I/\nu>100$ and Karlovitz number $Ka>1$. In the validation dataset $Re_t\approx200$ and $Re_T=u^{\prime}_{rms}l_I/(S_L l_F)\approx20$.

The standard, temperature based progress variable $c$ is defined as \cite{bray1985unified}
\begin{equation}
c=\frac{T-T_{u}}{T_{b}-T_{u}}
\end{equation}
Here, $T_u$ is the temperature of the unburnt reactants and $T_b$ is the temperature of the fully burnt products. Consider an iso-$c$ surface given by $c=c^{\star}$, within the flame structure of finite thickness. It is necessary to recognize that this surface could be wrinkled, multiply folded, disconnected and distributed over large part of the domain. The only necessary condition to define the iso-$c$ surface, locally, is $|\nabla c|\neq 0$. The mass flow rate passing through such an iso-$c$ surface is given by

\begin{equation}
\dot{m}_{c^{\star}}=-\int_{A_{T_{{c}^{\star}}}} \rho \vec{v}_{r} \cdot \hat{n} d A
\label{Eq: mass flow rate}
\end{equation}
Here $\rho$ is the local density, $\vec{v}_{r}$ is the flow velocity relative to the local surface, $\hat{n}$ is the local surface normal pointing towards lower temperature fluid and $A_{T_{{c}^{\star}}}$ is the total area of the surface defined by $c=c^{\star}$. Now, the local flow velocity relative to the local velocity of the surface is given by $\vec{v}_{r}$, where

\begin{equation}
\vec{v}_{r}=\vec{u}-\vec{v}_{f}
\label{Eq: relative velocity}
\end{equation}

Here $\vec{u}$ is the local flow velocity and $\vec{v}_{f}$ is the local surface velocity. The following identity for the local flame surface velocity is well known \cite{pope1988evolution}

\begin{equation}
\vec{v}_{f}=\vec{u}+S_{d} \hat{n}
\label{Eq: vf}
\end{equation}
where $S_{d}$ is the local flame displacement speed. Using Eqs. \ref{Eq: relative velocity} and \ref{Eq: vf} in Eq. \ref{Eq: mass flow rate}, we rewrite the mass flow rate through the surface as:

\begin{equation}
\dot{m}_{{c}^{\star}}=-\int_{A_ {T_{c^{\star}}}} \rho(-S_d \hat{n}) \cdot \hat{n} d A
\label{Eq: mdot Sd}
\end{equation}
We can define a surface specific turbulent flame speed $S_{T_{c^{\star}}}$ as
\begin{equation}
\rho_{u} S_{T_{c^{\star}}} A_{0}=\dot{m}_{{c}^{\star}}
\label{Eq: ST c=c*}
\end{equation}
$A_0$ is the area projected by $A_ {T_{c^{\star}}}$ onto the inlet plane of the cuboid. This is also of course equal to the constant cross sectional area of the cuboid.
Equating Eq. \ref{Eq: mdot Sd} and Eq. \ref{Eq: ST c=c*} we get
\begin{equation}
\rho_{u} S_{T_{{c}^{\star}}} A_{0}=-\int_{A_{T_{{c}^{\star}}}} \rho \left(-S_{d}\right) d A
\end{equation}
which yields
\begin{equation}
S_{T_{c^{\star}}}=\frac{1}{A_{0}} \int_{A_{T_{{c}^{\star}}}} \tilde{S_d} d A
\label{Eq: ST_c*_integral}
\end{equation}
Here $\tilde{S_d}=\rho S_d / \rho_u$ is the density weighted flame displacement speed. However, as noted before, the left hand side (LHS) represents a surface specific turbulent flame speed. We can transform it to a generalized turbulent flame speed by averaging over all $c=c^{\star}$, by integrating from $c_{min}$ to $c_{max}$. These are 0 and 1, respectively. Furthermore, we can average over a time period of $\tau=30\tau_I$ (as in the validation dataset) where $\tau_I$ is the integral eddy turnover time. Then, such a turbulent flame speed $S_T$ is defined as:

\begin{equation}
S_{T}=\frac{1}{\left(c_{\max }-c_{\min}\right) \tau} \int_{0}^{\tau}\int_{c_{\min}}^{c_{\max}} S_{T_{c^{\star}}} d c^{\star}dt
\label{Eq: ST_integral}
\end{equation}
Substituting Eq. \ref{Eq: ST_c*_integral} into Eq. \ref{Eq: ST_integral} we obtain:
\begin{equation}
S_{T}=\frac{1}{A_{0} \tau} \int_{0}^{\tau}\int_{0}^{1} \int_{A_{T_{c^{\star}}}}\tilde{S_{d}} d A d c^{\star} dt
\label{Eq: ST_formulation}
\end{equation}
\begin{figure}[h!]
\centering\includegraphics[trim=0cm 0cm 0cm 0cm,clip,width=1.0\textwidth]{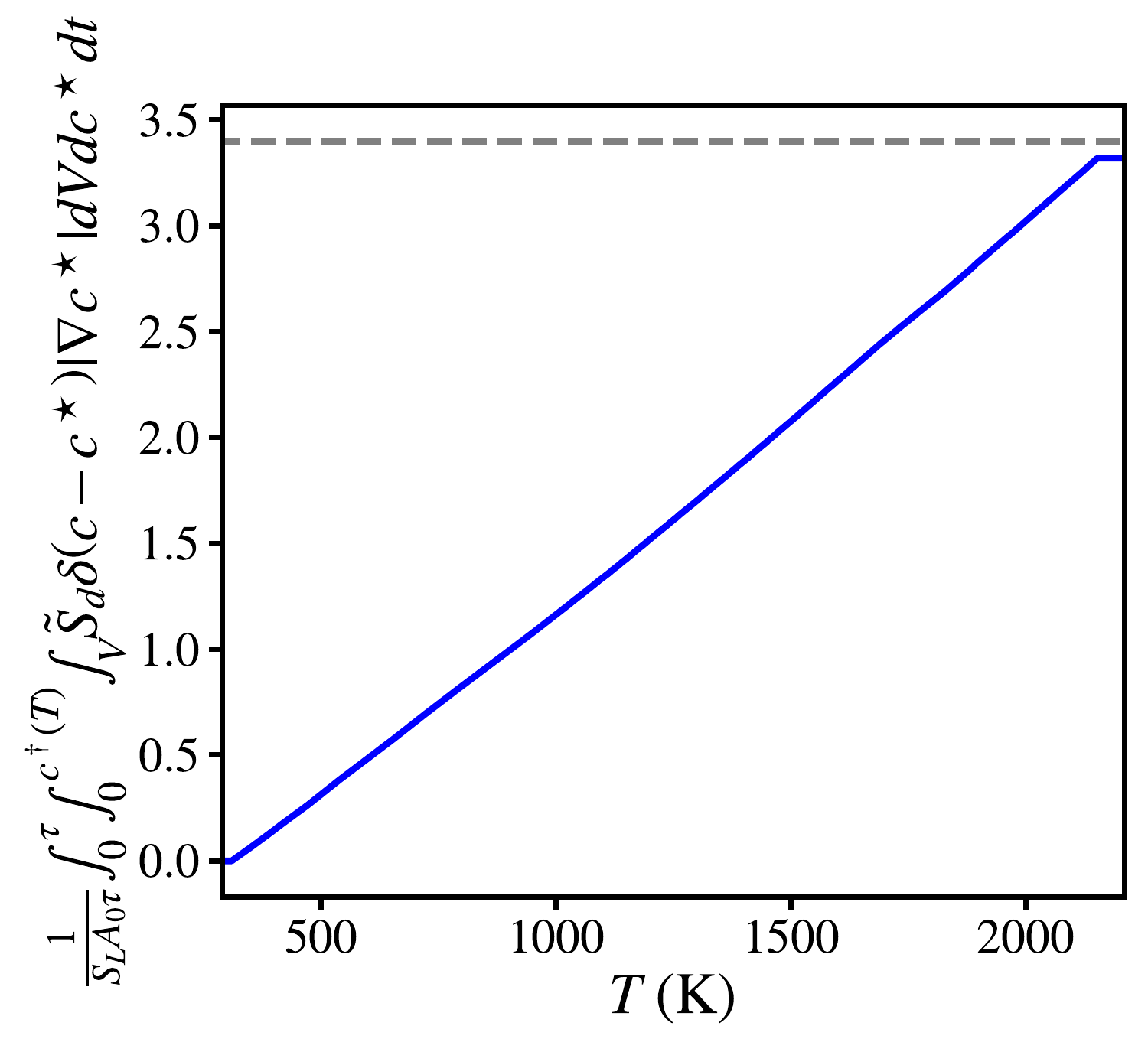}
\vspace{10 pt}
\caption{Convergence of $S_T$ defined using Eq. \ref{Eq: ST volume integral} onto global consumption speed $S_{T,GC}$ shown by the horizontal, dashed line near the top of the figure. Data plotted from the DNS dataset.}
\label{fig:integral Sd gradc deltac}
\end{figure}
Now, the area integral could be transformed to a volume integral using the delta function $\delta(c-c^{\star})$, \cite{osher2003level},
\begin{equation}
S_{T}=\frac{1}{A_{0} \tau} \int_{0}^{\tau} \int_{0}^{1} \int_{V} \tilde{S_d} \delta\left(c-c^{\star}\right)\left|\nabla c^{\star}\right| d V d c^{\star} dt
\label{Eq: ST volume integral}
\end{equation}
The integrand over $c^{\star}$ represents the time averaged mass flow rate through an isotherm, which should be nearly equal across all isotherms. This is reflected in the nearly linear plot shown in Fig. \ref{fig:integral Sd gradc deltac} obtained from the DNS dataset. The same figure also shows that the $S_T$ thus defined, does converge to the global consumption speed $S_{T,GC}$ as well. The average mass flow rate will indeed be equal to the global consumption rate of the reactants when all reactants are fully burned, as is the case in the present DNS.

Now, using the well-known relation
\begin{equation}
\int_{c^{\star}} f\left(c^{\star}\right) \delta\left(c-c^{\star}\right) d c^{\star}=f(c)
\end{equation}
we arrive at
\begin{equation}
S_{T}=\frac{1}{A_{0} \tau} \int_{0}^{\tau}\int_{V}{\tilde S_{d}}|\nabla c| d V dt
\label{Eq: ST simplified}
\end{equation}
Note that using the level set formulation given by the $G-$equation and starting with the initial condition $G(x,0)=x$, Kerstein et al. \cite{kerstein1988} obtained an expression similar to Eq. \ref{Eq: ST simplified} albeit in terms of gradient of $G$, without the time averaging. 
\begin{figure}[h!]
\centering\includegraphics[trim=0cm 0cm 0cm 0cm,clip,width=1.0\textwidth]{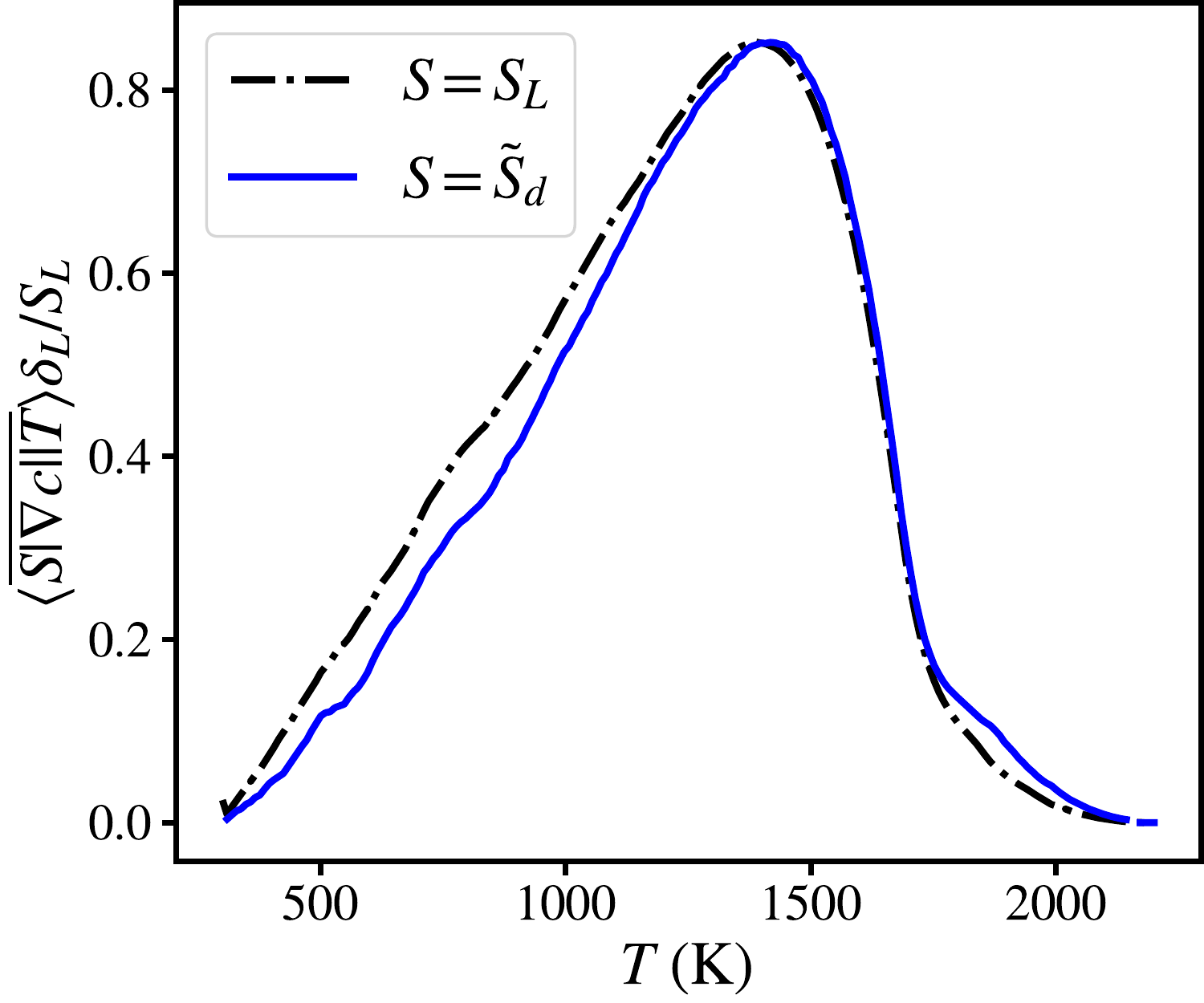}
\vspace{10 pt}
\caption{Approximating $\tilde{S_d}$ with $S_L$ in Eq. \ref{Eq: ST_volume_average}. Data plotted from the DNS dataset.}
\label{fig:Sd_SL_gradc_equivalence}
\end{figure}
Now, our volume of interrogation $V$ is always a fraction of the cuboid with $V$ having lateral dimensions $L$ and axial dimension equal to the mean flame brush thickness $\delta_T$, thereby just circumscribing the flame at all times. Also note $L^2=A_0$. Denoting time and volume averaging, with overbar and angular brackets respectively, Eq. \ref{Eq: ST simplified} yields:
\begin{equation}
S_{T}=\frac{L^{2} \delta_T}{A_{0}}\left\langle \overline{\tilde{S_d}|\nabla c|}\right\rangle
\label{Eq: ST_volume_average}
\end{equation}
The absence of subscripts in the angular brackets $\left\langle \right\rangle$ implies volume averaging whereas the subscript $y,z$ would imply planar averaging performed over the $y,z$ plane.
The RHS of this equation can be simplified with the assumption $\left\langle \overline{\tilde{S_d}|\nabla c|}\right\rangle \approx S_L \left\langle\overline{|\nabla c|}\right\rangle$ for unity Lewis number ($Le$) flames. For unity Lewis number flames at large $Ka$, this is actually an excellent assumption as shown in Fig. \ref{fig:Sd_SL_gradc_equivalence}. In the paragraphs below (till the end of the paragraph associated with Eq. \ref{Eq: STSL_delT_mod_gradc}) we show how a similar result could be obtained using closure models for $\tilde{S_d}$.

There is a vast literature establishing linear and non-linear relations between flame speed and stretch rate for steady, laminar premixed flames \cite{markstein1964, matalon1982, pelce1982, wu1985, law2006}. In turbulent premixed flames, the $S_d$ could be significantly different from $S_L$ \cite{echekki1996, chen1998correlation,hawkes2005evaluation, chakraborty2007stretch}. It has been recently shown that in turbulent premixed flames, the largest deviations of $\tilde{S_d}$ from $S_L$ originate from flame-flame interactions \cite{dave2020, yuvraj2022local} which ultimately leads to self-annihilation of the local flame surface \cite{trivedi2019topology}. Premixed flame-flame interaction has been an active research topic for the last two decades as exemplified in literature \cite{chen1999mechanism, griffiths2015three, talei2011sound, trivedi2019topology}. We can define $\kappa$ as the local curvature i.e. $\kappa = (\kappa_1 + \kappa_2)$ where $\kappa_1$ and $\kappa_2$ are the principal curvatures, with the convention that a surface convex towards reactants implies positive $\kappa$. Also, the thermal thickness of a standard laminar premixed flame is defined as $\delta_L=(T_b-T_u)/|\nabla T|_{max}$. In moderate to intense turbulence, portions of any iso-scalar surface within a flame are characterized by $\kappa \delta_L \leq -1$. Since their local radius of curvature is smaller than the flame thickness and the surface is propagating inwards, flame-flame interaction at some level is inevitable, in those portions. Since these portions of the flame are undergoing flame-flame interaction -- an inherently transient phenomenon leading to distortion of the flame structure itself -- the weak-stretch theories developed for stretched, steady, laminar premixed flames are not expected to be applicable. This was explicitly showed by Dave and Chaudhuri \cite{dave2020} who also analyzed a cylindrical, imploding, interacting and hence unsteady, laminar premixed flame, and showed that for the interacting stage ${S_d} \approx -2 {\alpha} \kappa$ while $\kappa \delta_L \ll 0$. Now, for the non-interacting portions that are weakly stretched, the local flame displacement speed could be explained \cite{dave2020} by the two-Markstein length weak stretch theories \cite{bechtold2001, giannakopoulos2015_1}. However, in those portions, to a leading order $\tilde{S_d} \approx S_L$. Indeed, the stretch rate deviates $\tilde{S_d}$ from $S_L$ but that deviation is typically much smaller than $S_L$ itself. In such a scenario, a model equation applicable for both non-interacting and interacting portions could be written as $\tilde{S_d} = S_L -2 \tilde{\alpha}_{0} \kappa$ for $\kappa \leq 0$, where density weighted thermal diffusivity $\tilde{\alpha}_{0}=\rho_0 \alpha_0 / \rho_u$; $\rho_0$ and $\alpha_0$ are density and thermal diffusivity, respectively, conditioned on a specific isotherm $T=T_0$ and $\rho_u$ is the unburned gas density.
This was validated using a DNS database of lean $H_2-air$ premixed flames \cite{yuvraj2022local}. As shown in Fig. \ref{fig: JPDFs_Sd_kappa}, the interacting flame model $\tilde{S_d} = S_L -2 \tilde{\alpha}_{0} \kappa$ do follow the overall trends of the JPDFs rather well, even for a n-heptane/air, high Karlovitz number premixed flame, over different isotherms. 
It appears from DNS data, that for $\kappa>0$, while a linear relation between $\tilde{S_d}-\kappa$ holds, the slope is different from $-2 \tilde{\alpha}_{0}$. As such Peters \cite{Peters2000} proposed a similar linear relation between $S_d$ and $\kappa$ across corrugated flamelet and thin reaction zone regimes, albeit with a different numerical coefficient. In any case, it will be shown in the following that a very specific form of the $\tilde{S_d}-\kappa$ relation is not necessary to arrive at a leading order scaling of $S_T$. Hence generic linear relationships between $\tilde{S_d}$ and $\kappa$ are considered:
\begin{figure}
     \centering
     \begin{subfigure}[b]{0.3\textwidth}
         \centering
         \includegraphics[width=\textwidth]{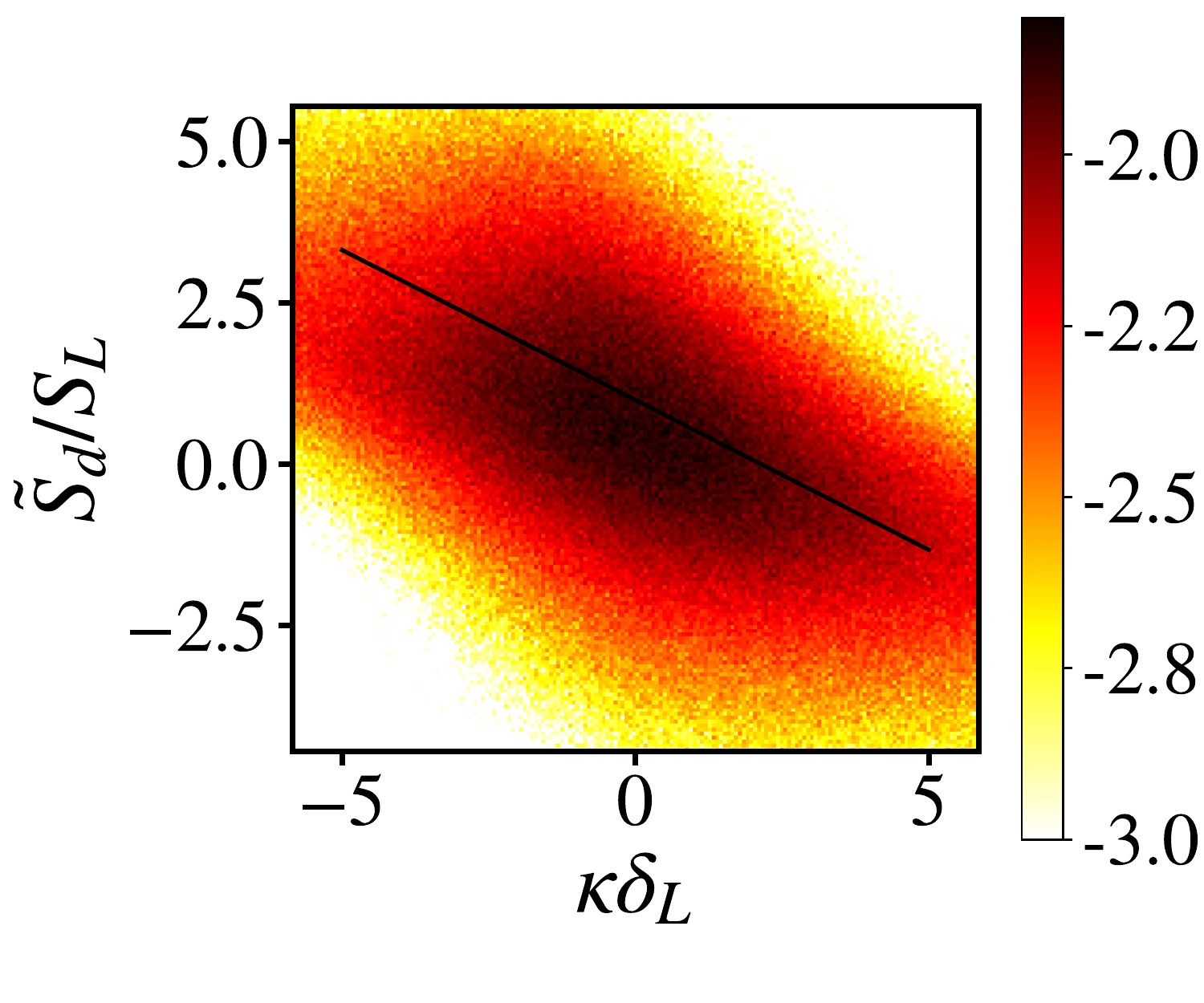}
         \caption{$T_0=500K$}
         \label{fig: 500K}
     \end{subfigure}
     \hfill
     \begin{subfigure}[b]{0.3\textwidth}
         \centering
         \includegraphics[width=\textwidth]{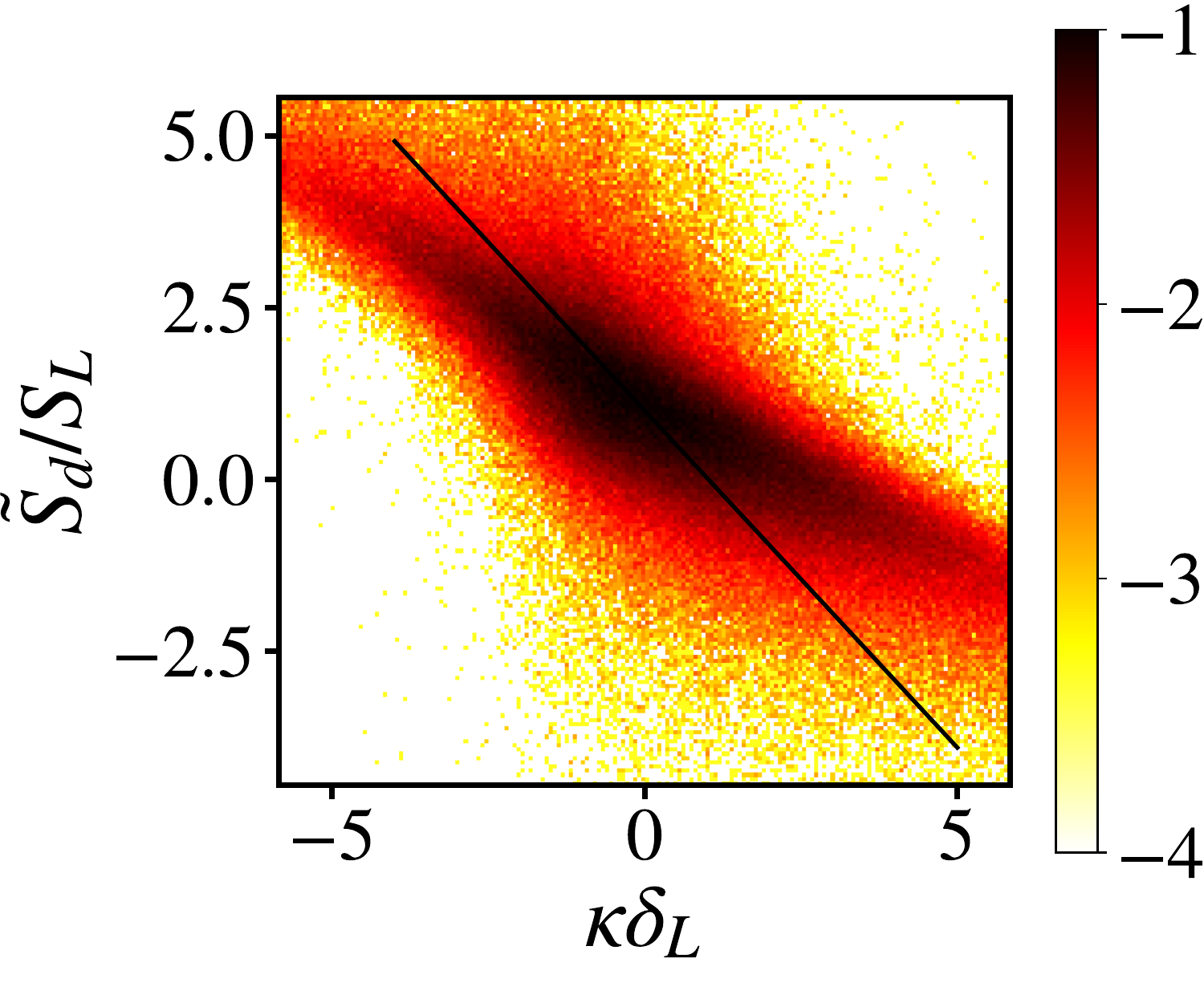}
         \caption{$T_0=1400K$}
         \label{fig: 1400K}
     \end{subfigure}
     \hfill
     \begin{subfigure}[b]{0.3\textwidth}
         \centering
         \includegraphics[width=\textwidth]{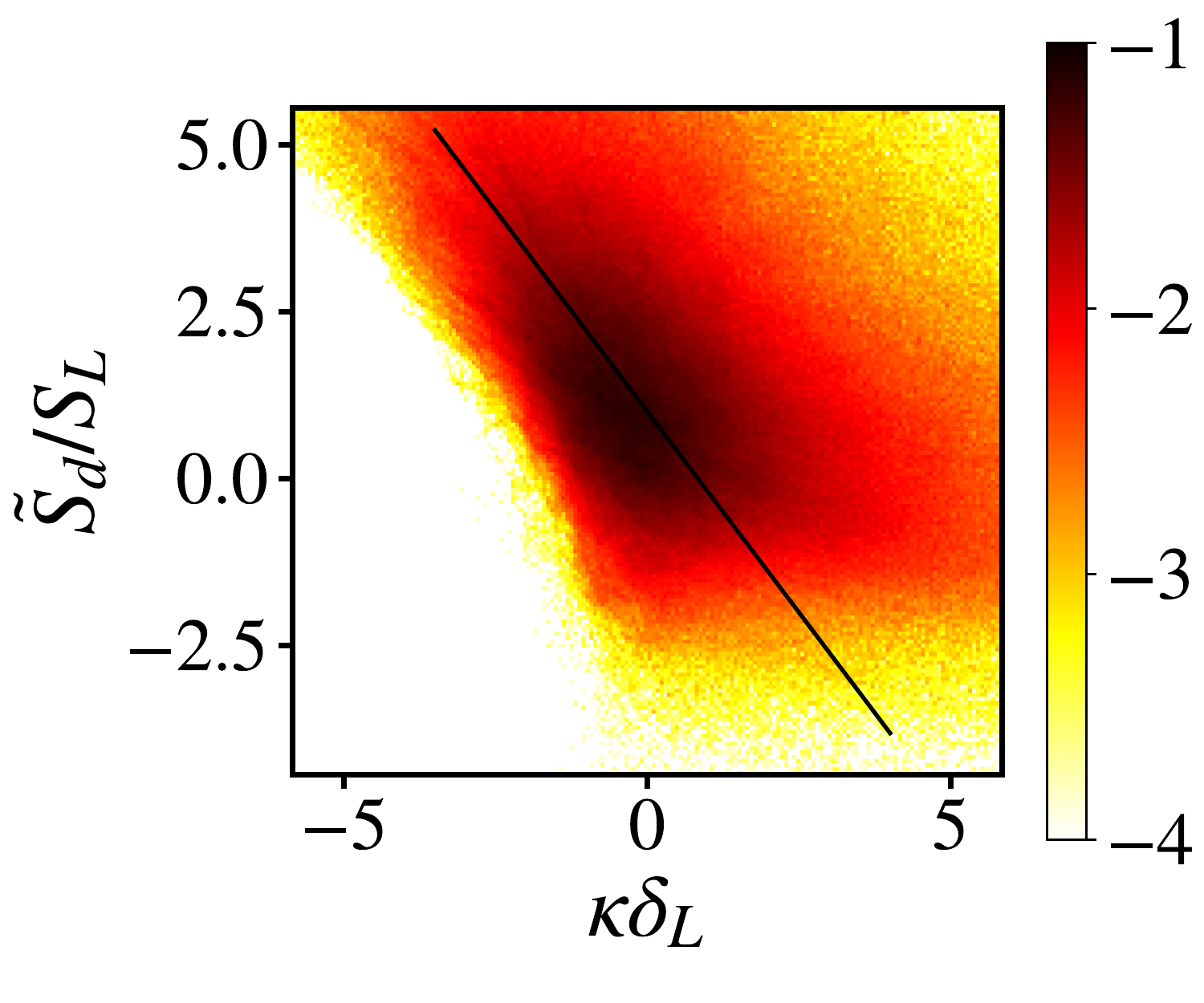}
         \caption{$T_0=1900K$}
         \label{fig: 1800K}
     \end{subfigure}
        \caption{Joint Probability Density Functions of normalized $\tilde{S_d}$ with normalized $\kappa$ as obtained from the DNS dataset. The colorbar shows the logarithm of the JPDF values. The black lines show the model $\tilde{S_d}=S_L - 2 \tilde{\alpha_0} \kappa$ }
        \label{fig: JPDFs_Sd_kappa}
\end{figure}
\begin{equation}
\begin{split}
\tilde{S_d} =S_{L} - \mathcal{A}_1 \kappa  \ \ \forall \kappa \leq 0 \\
\tilde{S_d} =S_{L} - \mathcal{A}_2 \kappa  \ \ \forall \kappa > 0
\end{split}
\label{Eq: Sd tilde}
\end{equation}
Note that $\mathcal{A}_1, \mathcal{A}_2$ are functions of $\tilde{\alpha}_{0}$, heat release rate, and $Le$ for non-unity $Le$ flames. For flames with $Le$ significantly lower than unity, these coefficients can be further influenced by differential diffusion and thermo-diffusive instability even in moderately large $Ka$ flames, as can be inferred from a recent work \cite{berger2022synergistic}. Substituting Eq. \ref{Eq: Sd tilde} into Eq. \ref{Eq: ST_volume_average} we get
\begin{equation}
S_{T}=\delta_T \left\langle S_{L} \overline{|\nabla c|}\right\rangle - {\delta_T \left\langle \mathcal{A}_1 \overline{\kappa |\nabla c|}\right\rangle}_{\kappa\leq0} - {\delta_T \left\langle \mathcal{A}_2 \overline{ \kappa |\nabla c|}\right\rangle}_{\kappa>0}
\label{Eq: ST_SL_expanded 2}
\end{equation}
Now, from DNS data it appears that for $|\kappa|\delta_L >1$, $\kappa |\nabla c| \approx k_0 $ holds, statistically. Here $k_0$ is an $\mathcal{O}(1)$ constant whose sign and magnitude could depend on the sign of $\kappa$. This can be explained from the limiting condition of flame-flame interaction which leads to very large negative $\kappa$, small $|\nabla c|$,  and large $\tilde{S_d}$. During the end stages of flame-flame interaction, as $\kappa \rightarrow \infty$, $|\nabla c| \rightarrow 0$. Furthermore, large positive $\kappa$ is typically generated by positive normal straining \cite{pope1988evolution}. Orthogonal to the positive normal strain rate, compressive strain rate aligns with the direction of the local scalar gradient in small $Da$ (and hence typically larger than unity $Ka$) turbulent premixed flames \cite{chakraborty2007influence} as in passive scalar turbulence \cite{ashurst1987alignment}, statistically. Clearly, extensive normal strain would lead to reduction of the magnitude of the scalar gradient while increasing positive curvature. Simultaneously, compressive normal strain would result in amplification  of the scalar gradient magnitude while reducing the positive curvature. Thus, increase (decrease) of $\kappa$ is often associated with decrease (increase) of $|\nabla c|$; hence their product could be assumed to be a constant. As such, from DNS data \cite{song2020dns, yuvraj2022local} it can be ascertained that for either sign of $\kappa$, the product $\kappa |\nabla c|$ tends to respective constants, statistically. We can write that for $\kappa \leq 0$, $\left\langle \overline{\kappa |\nabla c|}\right\rangle \approx -k_1$ while for $\kappa > 0$, $\left\langle \overline{\kappa |\nabla c|}\right\rangle \approx k_2.$ where both $k_1, k_2$ are positive constants. Substituting these into Eq. \ref{Eq: ST_SL_expanded 2} we get:
\begin{equation}
\frac{S_{T}}{S_{L}}=\delta_{T}\left\langle \overline{|\nabla c|} \right\rangle + \frac{\delta_T}{S_L} \left( \mathcal{A}_1 k_1 -  \mathcal{A}_2 k_2 \right)
\label{Eq: STSL_delT_mod_gradc}
\end{equation}
For close to unity or unity $Le$ flames, the second term is essentially a difference of two constants of similar magnitudes. The second term thus should be much smaller than the first term, and hence could be neglected. This essentially allowed introduction of $S_L$ in Eq. \ref{Eq: ST_volume_average}, reflected in Fig. \ref{fig:Sd_SL_gradc_equivalence}. For $Le<1$ ($Le>1$) we expect $\mathcal{A}_2 < \mathcal{A}_1$ ($\mathcal{A}_2 > \mathcal{A}_1$) and hence $S_T/S_L$ will be amplified (reduced) w.r.t. unity $Le$ conditions, all other parameters held fixed. Such amplification of $S_T$, for $Le<1$ flames is discussed extensively in \cite{lipatnikov2005molecular} and supported by DNS~\cite{aspden2016}, while the reduction of $S_T$ for $Le>1$ has been shown with DNS~\cite{savard2015broken,lapointe2016,aspden2016}. In any case, it is apparent that while the first term on the RHS of Eq. \ref{Eq: STSL_delT_mod_gradc} is the leading order term, in certain conditions, the second term on the RHS could lead to non-negligible corrections. A complete quantitative understanding of the entire second term needs further work. In the rest of the paper, we will focus on the leading order first term on the RHS of Eq. \ref{Eq: STSL_delT_mod_gradc} where $\left\langle \overline{|\nabla c|} \right\rangle$ is the important factor.
\begin{figure}[h!]
\centering\includegraphics[trim=0cm 0cm 0cm 0cm,clip,width=1.0\textwidth]{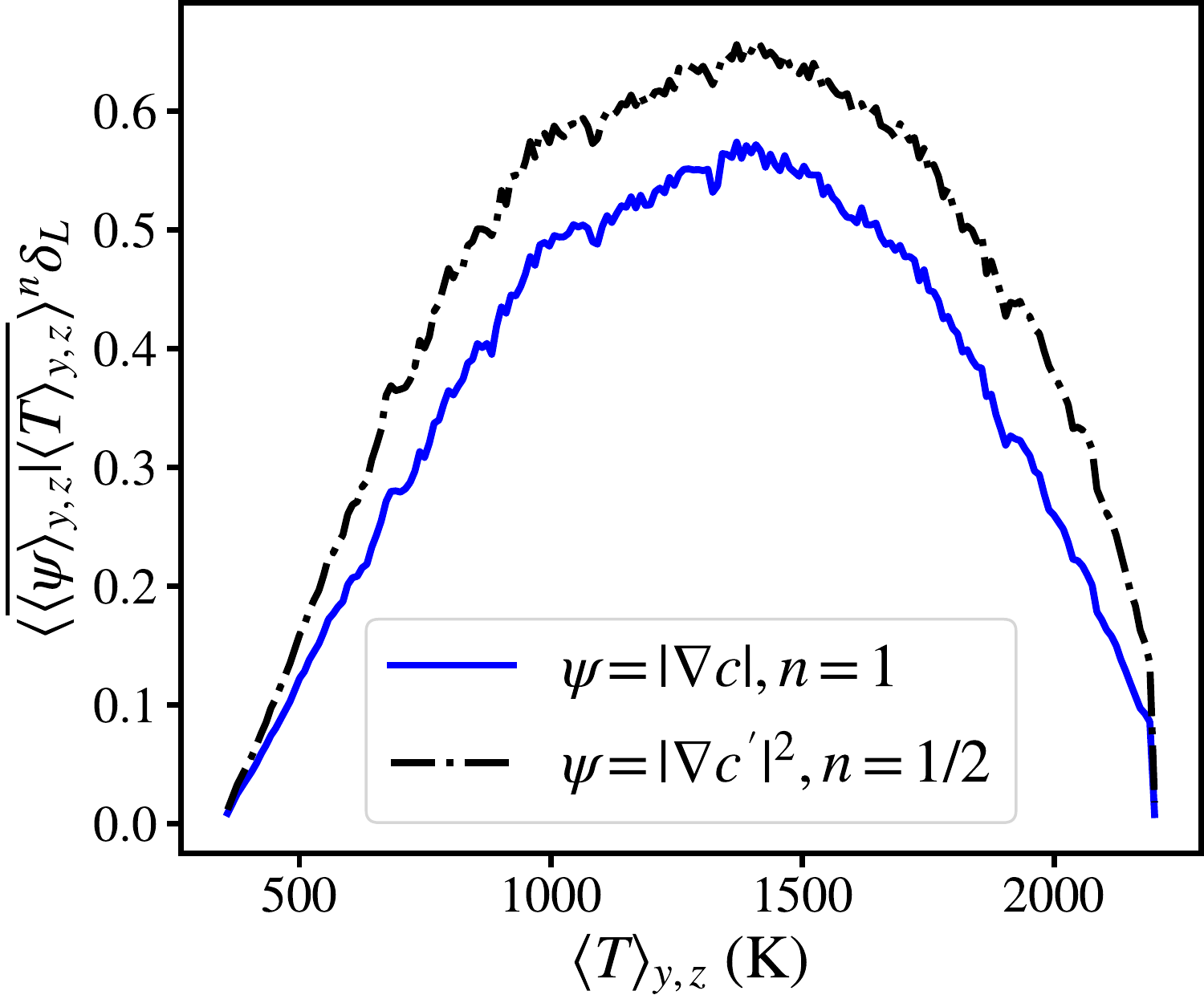}
\vspace{10 pt}
\caption{Comparing approximations used in Eq. \ref{Eq: grad c approx 1} and \ref{Eq: grad c approx 2}. Data from DNS dataset.}
\label{fig: grad c and cprime}
\end{figure}

Defining $c(x, y, z, t)=\left \langle{c}(x,y,z, t)\right\rangle_{y,z}+c^{\prime}(x, y, z, t)$. Note that $\left \langle \right \rangle _{y,z}$  represents averaging in the $y, z$ directions, i.e. average over the $y,z$ plane, at a fixed $x$ location. In intense turbulence, since statistically $ {|\nabla \langle c \rangle|} \ll  \left\langle {|\nabla c^{\prime}|}\right\rangle$, it is reasonable to assume:
\begin{equation}
\left \langle \overline {|\nabla c|} \right \rangle \approx  \left\langle \overline {|\nabla c^{\prime}|}\right\rangle
\label{Eq: grad c approx 1}
\end{equation}
Given that local scalar dissipation rate $\chi$ is nearly log-normally distributed, in intense turbulent premixed flames \cite{hamlington2012intermittency, chaudhuri2017flame},
it is also reasonable to assume $\left\langle\chi^n \right\rangle/\left\langle\chi \right\rangle ^n \approx 1$ where $\chi=2 \alpha  (\nabla c^{\prime} \cdot \nabla c^{\prime})$. Therefore, 
\begin{equation}
\left\langle \overline {|\nabla c|}
\right\rangle\approx\left\langle \overline {|\nabla c^{\prime}|} \right\rangle=\left\langle \overline {\left(\nabla c^{\prime} \cdot \nabla c^{\prime}\right)^{1 / 2}}\right\rangle \approx \left\langle \overline {\nabla c^{\prime} \cdot \nabla c^{\prime}}\right \rangle ^{1/2}
\label{Eq: grad c approx 2}
\end{equation}
These approximations are validated by comparing averages over $y,z$ planes, and over time, at fixed $x$ locations in Fig. \ref{fig: grad c and cprime}. Volume averages must be in close agreement given the similar variation of the planar averages.

\begin{figure}[h!]
\centering\includegraphics[trim=0cm 0cm 0cm 0cm,clip,width=1.0\textwidth]{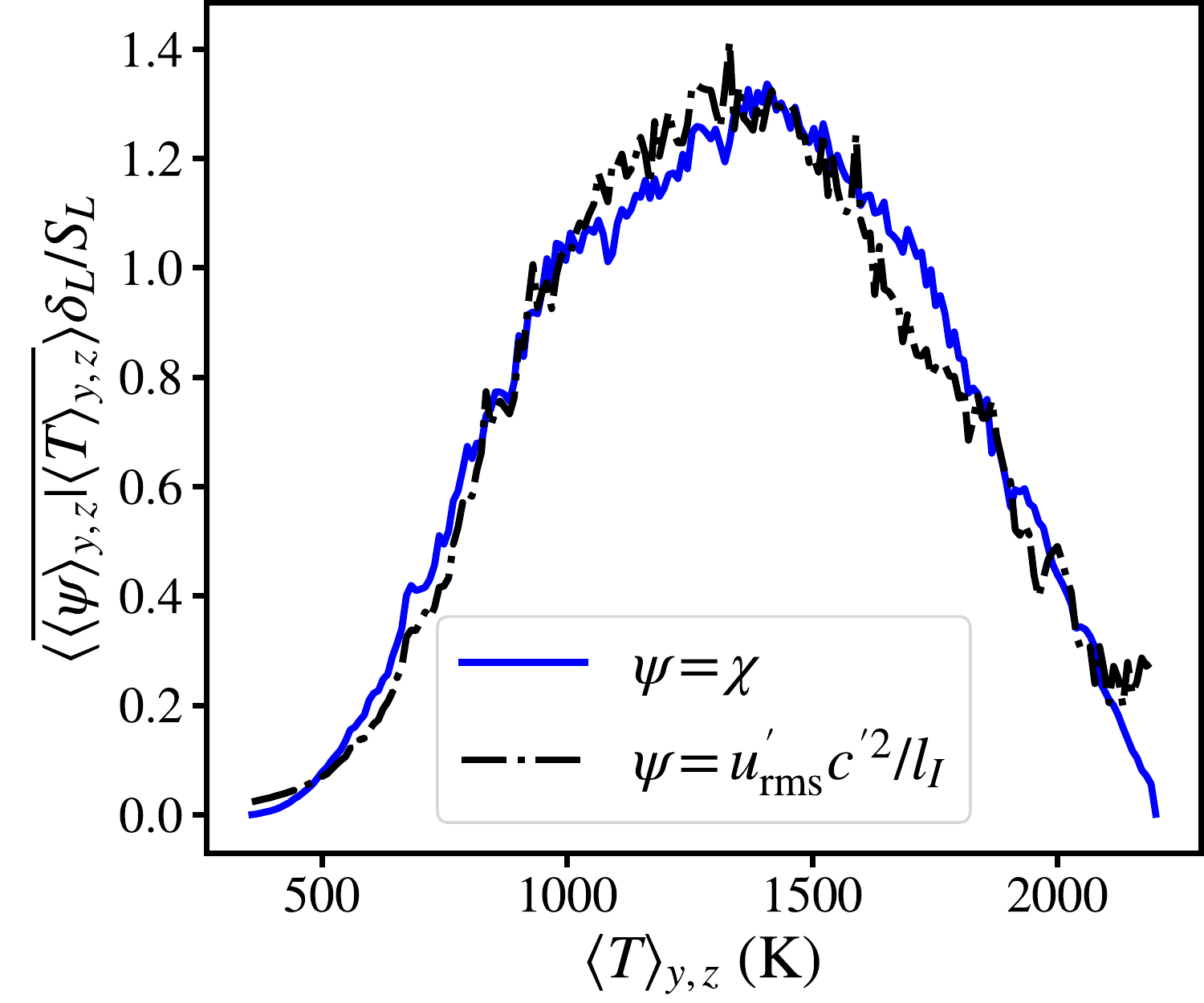}
\vspace{10 pt}
\caption{Comparison of time and $y-z$ plane averaged scalar dissipation rate $\left \langle \bar{\chi} \right \rangle_{y,z}$ with $u^{\prime}_{rms} \left \langle\overline{c^{\prime 2}}\right \rangle_{y,z}/ l_I$, justifying the approximation in Eq. \ref{mean SDR}. Data from DNS dataset.}.
\label{fig: SDR anamoly}
\end{figure}

Given that quasi-lognormal distribution of local scalar dissipation rate holds in intensely turbulent premixed flames, one can invoke the scaling originating from scalar dissipation rate anomaly (since the RHS in Eq. \ref{mean SDR} is independent of diffusivity, it is ostensibly an ``anomaly") well established for passive scalars in isotropic turbulence \cite{donzis2005scalar}. The implicit assumption is that for $Ka \geq \mathcal{O}(1)$, the scalar dissipation rate statistics is independent of dilatation originating from local heat release rate, to the leading order. Hence, we write:
\begin{equation}
\left \langle \overline {\chi} \right \rangle=2 \left\langle \alpha \overline {\nabla c^{\prime} \cdot \nabla c^{\prime}}\right\rangle \approx 2   \alpha_c   \left\langle \overline { \nabla c^{\prime} \cdot \nabla c^{\prime}}\right\rangle  \sim \frac{u_{r m s}^{\prime}\left \langle \overline {c^{\prime 2}}\right\rangle}{l_{I}}=\frac{u_{r}^{\prime}\left \langle \overline {c^{\prime 2}}\right \rangle_r} {r}
\label{mean SDR}
\end{equation}
Indeed, invoking the scalar dissipation rate anomaly is a major assumption, even for high $Ka$ flames. Therefore, in Fig. \ref{fig: SDR anamoly} we compare the $y,z,t$ averages $\left \langle\bar{\chi} \right \rangle_{y,z}=2   \alpha_c   \left \langle \overline{\nabla c^{\prime} \cdot \nabla c^{\prime}} \right \rangle_{y,z}$ and $u_{r m s}^{\prime} \left \langle\overline{c^{\prime 2}}\right \rangle_{y,z}/l_{I}$ as a function of the  $y,z$ plane averaged $\left \langle{T} \right \rangle_{y,z}$. As shown in Fig. \ref{fig: SDR anamoly} comparison of the planar averages show excellent agreement ensuring validity of the equality of their volume averages used in Eqn. \ref{mean SDR}.
This yields:
\begin{equation}
\left\langle \overline {|\nabla c|}\right \rangle \approx\left\langle \overline {|\nabla c^{\prime}|}\right \rangle \sim\left[\frac{u_{r m s}^{\prime} \left\langle \overline {c^{\prime 2}}\right\rangle}{2 \alpha_c  l_{I}}\right]^{1/2}
\label{Eq: mod_grad_c}
\end{equation}
Here $\alpha_c$ is a constant thermal diffusivity extracted from within the flame near the cross-over temperature such that $\alpha_c = S_L \delta_L$ 
The length scale $r$ could be any length scale within the inertial range.
Note that there exist sophisticated models for mean scalar dissipation rate \cite{vervisch2004three,kolla2009scalar} for turbulent combustion. However, in $Ka>1$ turbulence, the scalar dissipation rate anomaly scaling is expected to hold to the leading order, as evident from the scaling by Kolla et. al. \cite{kolla2009scalar} and also as already shown in Fig. \ref{fig: SDR anamoly}.
Substituting Eq. \ref{Eq: mod_grad_c} into Eq. \ref{Eq: STSL_delT_mod_gradc}, we get to the leading order:
\begin{equation}
\frac{S_{T}}{S_{L}} \sim \delta_T\left[\frac{u_{r m s}^{\prime} \left\langle \overline {c^{\prime 2}}\right\rangle}{2 \alpha_c l_I}\right]^{1 / 2}
\label{Eq: STSLcsquared}
\end{equation}
Next, we need to find a scaling relation for $\left\langle \overline {c^{\prime 2}}\right\rangle$. To that end, we linearize the monotonically increasing function of $x$: $\left \langle{c(x,y,z,t)}\right \rangle_{y,z}$ within the flame brush. It is implicitly assumed here that the domain is sufficiently large such that the $y,z$ averaging always yields converged statistics. The mean flow is along $x-$coordinate. $x$ also represents the axial direction of the cuboid with the coordinate fixed on the mean flame structure such that $\left \langle{c}(0,y,z,t)\right \rangle_{y,z}=0$; $y, z$ representing the transverse directions of the cuboid. Note that the $\left \langle{c}\right \rangle_{y,z}$ is not necessarily a linear function with distance. However, since the deviation from linearity may not be significant, it is assumed here, to retain mathematical tractability of the analysis. A more complex function like the error function could be used as well. Outside the flame brush   $\left \langle{c(x,y,z,t)}\right \rangle_{y,z}$ becomes 0 and 1 on the unburnt and burnt sides, respectively.
Assuming that $\left \langle {c(x,y,z,t)}\right \rangle_{y,z}$ is statistically stationary and hence is a function of $x-$ direction only:
\begin{equation}
\begin{aligned}
& \ \ \left \langle{c(x,y,z,t)}\right \rangle_{y,z}=\frac{x}{\delta_{T}} \\
&c(x, y, z, t)=\left \langle{c(x,y,z,t)}\right \rangle_{y,z}+c^{\prime}(x, y, z, t) \\
&\Rightarrow c=\frac{x}{\delta_{T}}+c^{\prime}
\end{aligned}
\label{Eq: x_c}
\end{equation}
$\left \langle{c(x,y,z,t)}\right \rangle_{y,z}$ represents $c(x, y, z, t)$ averaged over $y,z$ direction at any instant $t=t$. Next, we set:
\begin{equation}
\begin{aligned}
& c=c^{\star} \\
& \text{Therefore} \ \ c^{\star}=\frac{x_{c^{\star}}}{\delta_{T}}+c^{\prime} \\
&\Rightarrow x_{c^{\star}}=\delta_{T}\left(c^{\star}-c^{\prime}\right)
\end{aligned}
\end{equation}
Here, ${x_{c^{\star}}}$ denotes the $x-$coordinates of the surface $c=c^{\star}$. Utilizing ${x_{c^{\star}}}$ we can obtain the statistic of the distance over which the surface $c=c^{\star}$ fluctuates given by the variance of ${x_{c^{\star}}}$, denoted by $\ell_{T_{c^{\star}}}^{2}$

\begin{equation}
\ell_{T_{c^{\star}}}^{2}=\frac{1}{L^{2}} \int_{y} \int_{z}\left(x_{c^{\star}}- \left \langle {x}_{c^{\star}} \right\rangle_{y,z} \right)^2 d y d z
\label{Eq: ltsquare}
\end{equation}
Substituting Eq. \ref{Eq: x_c} into Eq. \ref{Eq: ltsquare} we get:
\begin{equation}
\ell_{T_{c^{\star}}}^{2}=\frac{\delta^2_T}{L^{2}} \int_{y} \int_{z}\left(c^{\star} - c^{\prime} - c^{\star} + \left \langle {c^{\prime}} \right\rangle_{y,z}\right)^2 d y d z
\end{equation}
Since, $\left \langle {c^{\prime}} \right\rangle_{y,z}=0$ from Eq. \ref{Eq: x_c}, the above equation leads to
\begin{equation}
\ell_{T_{c^{\star}}}^{2}=\frac{\delta^2_T}{L^{2}} \int_{y} \int_{z}\left(- c^{\prime}\right)^2 d y d z
\end{equation}
Next we average the above equation over the flame brush thickness $\delta_T$ in the $x$ direction and over time to obtain a generalized flame surface fluctuation distance $\ell_{T}$ given by
\begin{equation}
\ell_{T}^{2}=\frac{\delta^2_T}{\tau L^{2} \delta_T} \int_{0}^{\tau} \int_{x} \int_{y} \int_{z}\left(- c^{\prime}\right)^2 d x d y d z dt
\end{equation}
This yields a closure for $\left\langle \overline {c^{\prime 2}}\right\rangle$ in terms of $\delta_T$ and $\ell_{T}$. 
\begin{equation}
\ell_{T}^{2}=\delta^2_T\left\langle \overline {c^{\prime 2}}\right\rangle
\label{Eq: cprime_square}
\end{equation}
\begin{figure}[h!]
\centering\includegraphics[trim=0cm 0cm 0cm 0cm,clip,width=1.0\textwidth]{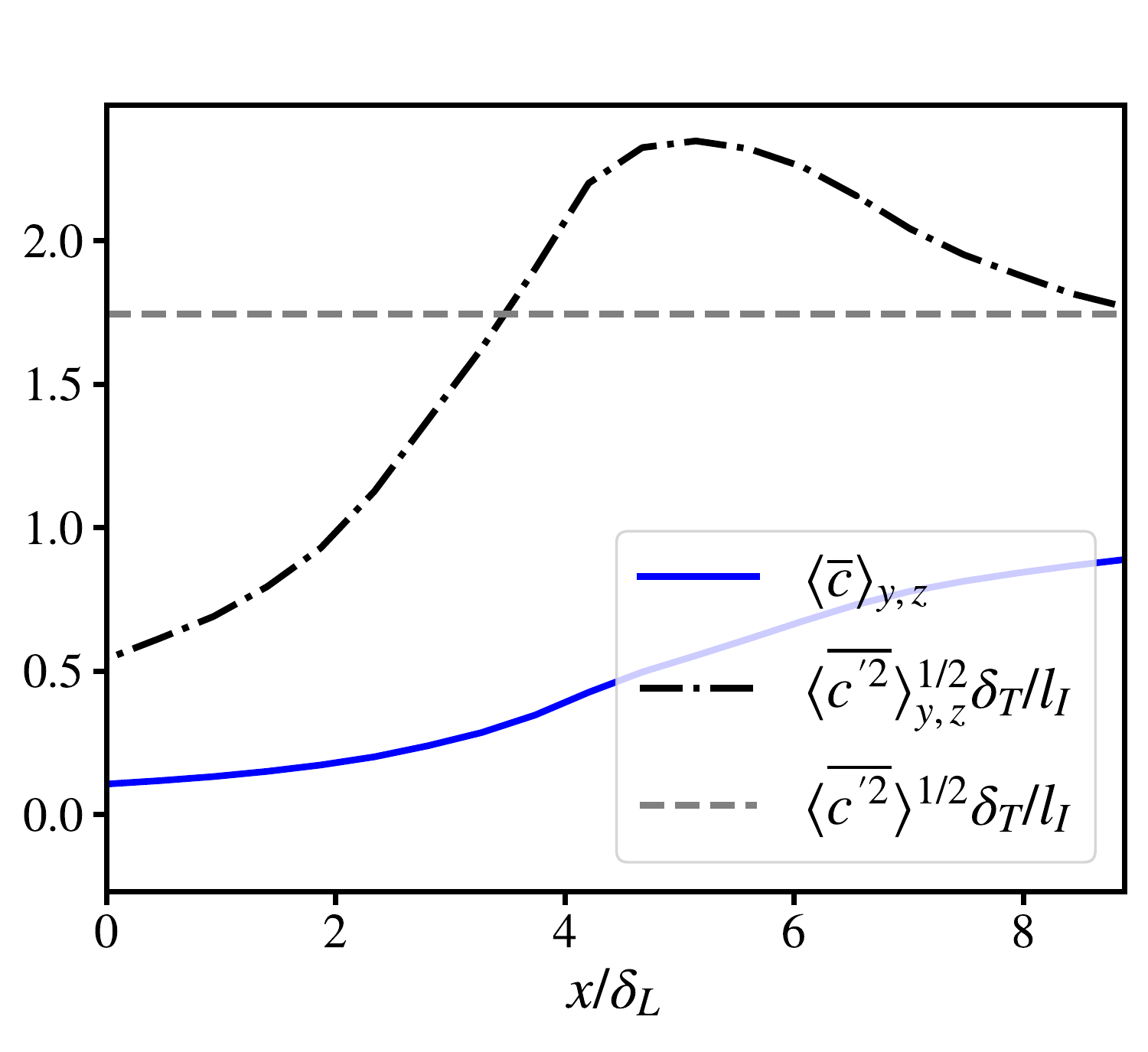}
\vspace{10 pt}
\caption{Plot showing variation of time and $y-z$ plane averaged progress variable $\left \langle \overline{c} \right \rangle_{y,z}$, $\left \langle \overline{ c^{\prime 2} }\right \rangle^{1/2}_{y,z} \delta_T/l_I$ vs $x/\delta_L$. The horizontal dashed line representing time and volume averaged $\left \langle \overline{ c^{\prime 2} }\right \rangle^{1/2}$ shows that $\ell_T=\delta_T\left\langle \overline {c^{\prime 2}}\right\rangle^{1/2} =1.75l_I$.}
\label{fig: fluctuation distance assuption}
\end{figure}
Substituting Eq. \ref{Eq: cprime_square} into Eq. \ref{Eq: STSLcsquared} we get
\begin{equation}
\frac{S_{T}}{S_{L}} \sim \left[\frac{u_{r m s }^{\prime}  \ell_{T}^2 }{S_L \delta_L l_{I}}\right]^{1/2}
\end{equation}
Here we used $ \alpha_c  = S_L \delta_L$ where $\delta_L = (T_b - T_u)/ |\nabla T|_{max}$ i.e. the thermal thickness of the standard laminar premixed flame. Note that $\delta_L$ is different from
$l_F$ which is the diffusion thickness of a standard laminar premixed flame. The largest flame surface fluctuations determining its variance should be induced by the largest length scales of turbulence. Hence, it is reasonable to assume $\ell_{T} \sim l_I$, yielding
\begin{equation}
\frac{S_{T}}{S_{L}} \sim \left[\frac{u_{r m s }^{\prime}  l_I }{S_L \delta_L}\right]^{1/2}
\end{equation}
Assuming $ \alpha_c \approx \nu_c $, we get
\begin{equation}
\frac{S_{T}}{S_{L}} \sim Re_T^{1/2}
\label{STSL scaling}
\end{equation}
to the leading order. The assumption that $\ell_T=\delta_T\left\langle \overline {c^{\prime 2}}\right\rangle^{1/2} \sim l_I$ (where $\ell_T$ is defined by Eq. \ref{Eq: cprime_square}) is justified by Fig. \ref{fig: fluctuation distance assuption} which shows $\ell_T=1.75l_I$. 

The scaling given by Eqn. \ref{STSL scaling} is also the expression that was obtained by Damk{\"o}hler \cite{damkohler1940einfluss} in the intense turbulence limit, but with limitations discussed in the Introduction. Equation \ref{STSL scaling} is also consistent, in part, with the results from \cite{Peters2000, kolla2009scalar, chaudhuri2011, sabelnikov2021scaling}, often derived for global consumption speed, using different techniques and associated assumptions.

The form of the scaling provided by Eq. \ref{STSL scaling} is consistent with the ``bending behavior" of turbulent flame speed observed in DNS. However, a very large database focusing on turbulent flame speed scaling over a wide range of $Ka$ is rare.  Using DNS with detailed chemistry, Aspden et al. \cite{aspden2019towards} showed that the consumption speed $S_{T,GC} \sim Re_T^{1/2}$ holds over three decades of $Ka$ out of the four decades of $Ka$ investigated for both $H_2-$ and $CH_4$-air turbulent premixed flames. However, the number of data-points are limited. When reactants are fully consumed the $S_T$ defined in the present paper should be equal to $S_{T,GC}$. The large DNS database of Yu and Lipatnikov \cite{yu2017dns} obtained using single step chemistry clearly showed that $S_{T,GC}/S_L \sim Re_T^{1/2}$, for small $Da$. Experimentally obtained statistically planar turbulent flame speeds are not common. The configuration that is closest to the statistically planar turbulent premixed flame is the turbulent expanding flame configuration. However, the results derived for statistically planar flame configuration cannot be directly applied in the expanding flame configuration.
For an expanding premixed flame with average radius $\langle R \rangle \gg \delta_L$, we can consider a sector of the turbulent expanding flame to be a statistically quasi-planar, quasi-steady flame enclosed in a cuboid box with lateral dimensions equal to the $\langle R \rangle$. Indeed the characteristic dimensions of the expanding flame $\langle R \rangle$, flame-brush thickness $\delta_T$ are monotonically increasing with time $t$. Furthermore, $\delta_T$ monotonically increases with $\langle R \rangle$. In view of this the averaged scalar dissipation rate, measured within the cuboid of lateral dimension $\langle R \rangle$ is obtained from Eq. \ref{mean SDR}, as:

\begin{equation}
\langle \chi \rangle_{\langle R \rangle}=2 \left\langle \alpha \nabla c^{\prime} \cdot \nabla c^{\prime}\right\rangle _{\langle R \rangle} \sim \frac{u_{{\langle R \rangle}}^{\prime}\left \langle c^{\prime 2}\right \rangle_{\langle R \rangle}} {{\langle R \rangle}}
\end{equation}
The rest of the derivations proceeds similar to that of the statistically planar flames with the difference that in the expanding flame configuration the flame surface fluctuation length scale $\ell_T \sim \langle R \rangle$. Substituting these yields:
\begin{equation}
\frac{S_{T}}{S_{L}} \sim \left[\frac{u_{\langle R \rangle}^{\prime}  \langle R \rangle }{S_L \delta_L}\right]^{1/2}
\end{equation}
Similar scaling has been experimentally demonstrated by at least five groups in the last decade \cite{chaudhuri2012flame, chaudhuri2013scaling, wu2015propagation, jiang2016high, zhao2021flame, fries2019flame, kulkarni2021reynolds}. Indeed, given the constitutive relations, scalings introduced, the match between the experimental and theoretical results should not be considered evidence of a ``proof" that $S_T/S_L \sim Re_T^{1/2}$ is universal. Different scalings resulting from physics or configuration not considered in the present derivation is certainly possible and merit further exploration.
\section{Conclusions}
In this paper, we revisit turbulent flame speed, from first principles. Starting with the mass flow rate through an ensemble of isotherms constituting a statistically planar turbulent premixed flame, a leading order scaling of the normalized turbulent flame speed: $S_T/S_L\sim Re_T^{1/2}$ is obtained. Major approximations and interim results are validated with the high $Ka$, unity $Le$, DNS of n-heptane and air mixture, computed with a reduced chemical reaction mechanism. The scalar dissipation rate anomaly is used alongside a linearized mean progress variable profile which yields a simplified scaling of the isotherm fluctuation distance -- a new length scale. Finally, the scaling obtained is extended to sufficiently large expanding flames experiencing negligible mean stretch rate. 
\section{Acknowledgements}
SC sincerely acknowledges the valuable comments by Prof. Andrei Lipatnikov, Mr. Yuvraj, and Mr. Yazdan Naderzadeh. This research was enabled in part by support provided by the Natural Sciences and Engineering Research Council of Canada through Discovery Grants (RGPIN-2021-02676 and RGPIN-2019-04309) and Heuckroth Distinguished Faculty Award in Aerospace Engineering from UTIAS. Computational resources for the DNS post-processing were provided by Compute Ontario and Compute Canada.

\bibliographystyle{ieeetr}
\bibliography{sample.bib}







\end{document}